\documentstyle[12pt]{article}
\begin{document}
\begin{flushright}
{\bf
hep-th/0001150 \\
2000, January 20}
\end{flushright}
\begin{center}

 {\Large\bf
 Super-D0-branes \\ at the endpoints of fundamental
 superstring: \\ an example of interacting brane system}
 \footnote{Contribution to the Proceedings of the
International Seminar ''Supersymmetries and Quantum Symmetries'
(SQS'99, 27-31 July, 1999).}

\bigskip

{\bf Igor BANDOS}

\bigskip

{\it Institute for Theoretical Physics, \\
NSC Kharkov Institute of Physics and Technology,\\
 310108, Kharkov,  Ukraine} \\
{e-mail: bandos@hep.itp.tuwien.ac.at \\
bandos@kipt.kharkov.ua}

\bigskip

{\bf Abstract}
\end{center}

{\small We present a supersymmetric action functional for the
coupled system of
an open fundamental superstring and super--D0--branes attached to
(identified with) the string endpoints.
As a preliminary step the geometrical actions
for a free super-D0-brane and a free type IIA superstring have
been built. The pure bosonic limits of the action for the coupled system
and of the equations of motion are discussed in some detail.}

\section*{Introduction}

Recently a way to obtain
a supersymmetric action functional for interacting branes
(intersecting branes and branes ending on branes)
has been proposed  \cite{BK}.
The systems involving open fundamental superstrings
ending on super--Dp--branes are quite  generic and, on the other hand,
especially interesting. The case of superstring---super-D3-brane system
has been discussed briefly in \cite{BK} (see \cite{BK2} for details).

Two types of such system are special and require separate consideration.
One consists of the open superstring and a super--D9--brane
(space--time filling brane). It has been elaborated in
\cite{BKD9}.
In this contribution we present a supersymmetric action functional
for the system of the open superstring ending
on (the dynamical) super-D0-branes or D-particles.
In distinction to the general case \cite{BK,BK2} neither
Lagrange multipliers no auxiliary space--time filling brane are necessary
in this case.

Note that this dynamical system provides a supersymmetric generalization
of the 'string with masses at the endpoints' which has been considered in
the early years of 'QCD string' \cite{BarNes}.

\section{Geometric action for free super-D0-brane}

The geometric action \cite{BZ} and the generalized action principle
\cite{bsv} for super--Dp--branes with $0< p< 9$ and $p=9$ has been
constructed in \cite{bst,abkz} respectively. However the super--D0--brane
has not been considered in this framework.

The geometric action for the super--D0--brane has the form
\begin{equation}\label{SD0}
S_{D0}=  m \int_{{\cal M}^1}
{\tilde{\cal L}}_1 =
\int_{{\cal M}^1}
\left(
\tilde{\Pi}^{\underline{m}} u^{(0)}_{\underline{m}}(\tau) +
i  \left( d\tilde{\Theta}^{1\underline{\mu}}
\tilde{\Theta}^{2}_{\underline{\mu}} -
\tilde{\Theta}^{1\underline{\mu}} d\tilde{\Theta}^{2}_{\underline{\mu}}
\right) \right),
\end{equation}
where  $m$ is the super-D0-brane mass parameter,
\begin{equation}\label{Pi}
{\Pi}^{\underline{m}} =
dX^{\underline{m}} -
i d{\Theta}^{1\underline{\mu}}
{\sigma}^{\underline{m}}_{\underline{\mu}\underline{\nu}}
{\Theta}^{1\underline{\nu}}
-
i d{\Theta}^2_{\underline{\mu}}
\tilde{\sigma}^{\underline{m} \underline{\mu}\underline{\nu}}
{\Theta}^2_{\underline{\nu}}
\end{equation}
is the basic covariant 1--form of the flat type IIA superspace,
\begin{equation}\label{tPi}
\tilde{\Pi}^{\underline{m}} =
d\tilde{X}^{\underline{m}} -
i d\tilde{\Theta}^{1}
{\sigma}^{\underline{m}}
\tilde{\Theta}^1
-
i d\tilde{\Theta}^2
\tilde{\sigma}^{\underline{m}}
\tilde{\Theta}^2 = d\tau \tilde{\Pi}^{\underline{m}}_\tau
\end{equation}
is its pull--back on the super--D0--brane world--line
${\cal M}^1$
\begin{equation}\label{M1}
{X}^{\underline{m}} =
\tilde{X}^{\underline{m}} (\tau), \quad
{\Theta}^{1\underline{\mu}} =
\tilde{\Theta}^{1\underline{\mu}}
 (\tau), \quad
{\Theta}^2_{\underline{\mu}} =
\tilde{\Theta}^2_{\underline{\mu}}
 (\tau) \quad : \quad {\cal M}^1 \rightarrow {\underline{{\cal
 M}}}^{(10|32)},
\end{equation}
and $u^{(0)}_{\underline{m}}$ is a time--like unit length vector
field
\begin{equation}\label{u0}
u^{(0)}_{\underline{m}} u^{(0)\underline{m}} = 1.
\end{equation}
It is convenient to consider $u^{(0)}_{\underline{m}}$ as a column
of the Lorentz group valued matrix
\begin{equation}\label{vharm}
u^{~\underline{a}}_{\underline{m}} =
\left( u^{0}_{\underline{m}}, u^{~i}_{\underline{m}}\right)~~ \in ~~
SO(1,9) \qquad \Leftrightarrow \qquad
u^{~\underline{a}}_{\underline{m}}
\eta^{\underline{m}\underline{n}} u^{~\underline{b}}_{\underline{n}} =
\eta^{\underline{a}\underline{b}}.
\end{equation}
The conditions (\ref{vharm}) include the normalization (\ref{u0}) as well
as the orthogonality conditions for the vectors
$
u^{(0)}_{\underline{m}},
u^{~i}_{\underline{m}}
$
(Lorentz harmonics \cite{Sok})

\begin{equation}\label{ui}
u^{(0)}_{\underline{m}} u^{i\underline{m}} = 0, \qquad
u^{i}_{\underline{m}} u^{j\underline{m}} = - \delta^{ij}.
\end{equation}
A doubly covered element for the $SO(1,9)$--valued matrix (\ref{vharm})
\begin{equation}\label{sharmD0}
v^{~A}_{\underline{\mu}} \quad \in \quad Spin(1,9)
\end{equation}
(spinor Lorentz harmonics, see \cite{BZ} and refs. therein)
is related with (\ref{vharm}) by the conditions of $\sigma$--matrix
conservation
\begin{equation}\label{vss}
 u^{~\underline{a}}_{\underline{m}}
\sigma^{\underline{m}}_{\underline{\mu}\underline{\nu}}
= v_{\underline{\mu}}^{~A}
\sigma^{\underline{a}}_{AB} v^{~B}_{\underline{\mu}},
\qquad
u^{~\underline{a}}_{\underline{m}}
\tilde{\sigma}_{\underline{a}}^{AB}
= v_{\underline{\mu}}^{~A}
\tilde{\sigma}_{\underline{m}}^{\underline{\mu }\underline{\nu }}
v^{~B} _{\underline{\mu}}.
\end{equation}
$A=1,\ldots , 16$ can be treated as $SO(9)$ spinor index. Then the
requirement of  $SO(9)$ gauge symmetry makes natural an identification of
the  harmonics with homogeneous coordinates of the
coset $SO(1,9)/SO(9)$ (cf.  with \cite{GIKOS}).
Note that $SO(9)$ group possesses a symmetric
charge conjugation matrix. When it is identified with unity matrix, the
difference between upper and lower $SO(9)$ spinor indices disappears.

Substituting  the $SO(9)$ invariant representation for $SO(1,9)$
sigma-matrices
\begin{equation}\label{sigma}
\sigma^{0}_{AB} = \delta_{AB}, \quad
\sigma^{i}_{AB} = \Gamma^i_{AB}, \quad
\tilde{\sigma}^{0~AB} = \delta_{AB}, \quad
\tilde{\sigma}^{i~AB} = - \Gamma^i_{AB}, \quad
\end{equation}
one can decompose  Eq. (\ref{vss}) into
\begin{equation}\label{vss1}
 u^{(0)}_{\underline{m}}
\sigma^{\underline{m}}_{\underline{\mu}\underline{\nu}}
= v_{\underline{\mu}}^{~A}
v^{~A}_{\underline{\nu}},
\qquad
 u^{i}_{\underline{m}}
\sigma^{\underline{m}}_{\underline{\mu}\underline{\nu}}
= v_{\underline{\mu}}^{~A} \Gamma^i_{AB}
v^{~B}_{\underline{\nu}}.
\end{equation}
\begin{equation}\label{vss2}
 u^{(0)}_{\underline{m}} \delta_{AB} =
 v_{\underline{\mu}}^{~A}
\tilde{\sigma}_{\underline{m}}^{\underline{\mu}\underline{\nu}}
v^{~B}_{\underline{\nu}},
\qquad
 u^{i}_{\underline{m}}
\Gamma^i_{AB}
= v_{\underline{\mu}}^{~A}
\tilde{\sigma}_{\underline{m}}^{\underline{\mu}\underline{\nu}}
v^{~B}_{\underline{\mu}}.
\end{equation}

 Similar relations can be obtained for the inverse $SO(1,9)/SO(9)$
 harmonics
\begin{equation}\label{s-1}
v_{B}^{~\underline{\mu}}
v_{\underline{\mu}}^{~A}
= \delta_{B}^{~A},
\qquad
\end{equation}
\begin{equation}\label{vss-1}
 u^{(0)}_{\underline{m}}
\tilde{\sigma}^{\underline{m}~\underline{\mu}\underline{\nu}}
= v^{~\underline{\mu}}_{A}
v_{A}^{~\underline{\nu}},
\qquad
 u^{i}_{\underline{m}}
\tilde{\sigma}^{\underline{m}~\underline{\mu}\underline{\nu}}
= - v^{~\underline{\mu}}_{A} \Gamma^i_{AB}
v_{A}^{~\underline{\nu}}.
\end{equation}
\begin{equation}\label{vss-2}
 u^{(0)}_{\underline{m}} \delta_{AB} =
 v^{~\underline{\mu}}_{A}
\sigma_{\underline{m}~\underline{\mu}\underline{\nu}}
v_{B}^{~\underline{\nu}},
\qquad
 u^{i}_{\underline{m}}
\Gamma^i_{AB}
= v^{~\underline{\mu}}_{A}
\sigma_{\underline{m}\underline{\mu}\underline{\nu}}
v_{ B}^{~\underline{\mu}}.
\end{equation}

 The harmonics can be used to define a general supervielbein of the flat
 type $IIA$ superspace
 $E^{{\cal A}}$ which possesses the $SO(9)$ invariant decomposition
 $E^{{\cal A}} = (E^{(0)}, E^i; E^{A1}, E^{A2})$
 \begin{equation}\label{E0i}
 E^{(0)} \equiv \Pi^{\underline{m}}
 u^{(0)}_{\underline{m}}, \qquad E^i \equiv \Pi^{\underline{m}}
 u^{i}_{\underline{m}} , \end{equation} \begin{equation}\label{EA1,2}
 E^{A1} \equiv d\Theta^{\underline{\mu}1} v^{~A}_{\underline{\mu}}, \qquad
 E^{A2} \equiv d\Theta^2_{\underline{\mu}} v_{A}^{~\underline{\mu}}.
\qquad
\end{equation}
A new important property of this supervielbein (in comparison with
the ''coordinate'' one $(\Pi^{\underline{m}}, d\Theta^{\underline{\mu}1},
d\Theta^2_{\underline{\mu}})$) is that it permits covariant linear
combinations of the different fermionic supervielbein forms, e.g.
$E^{A1} \pm E^{A2}$.

The structure equations of the flat type $IIA$ superspace
can be written as
\begin{equation}\label{dE0} dE^{(0)} = -i  E^{A1} \wedge E^{A1} -i
E^{A2} \wedge E^{A2} + E^i \wedge f^i , \qquad \end{equation}
\begin{equation}\label{dEi}
{\cal D}E^{i} \equiv  dE^i + E^j \wedge A^{ji} =
-i  E^{A1} \wedge E^{B2} \Gamma^i_{AB} + E^{(0)} \wedge f^i , \qquad
\end{equation}
\begin{equation}\label{dEA1}
{\cal D} E^{A1} \equiv  dE^{A1} + E^{B1} \wedge
{1 \over 4} A^{ij}
\Gamma^{ij}_{BA} =
{1 \over 2} E^{B1} \wedge
f^{i}
\Gamma^{i}_{BA}, \qquad
\end{equation}
\begin{equation}\label{dEA2}
{\cal D}E^{A2} \equiv  dE^{A2} + E^{B2} \wedge
{1 \over 4} A^{ij}
\Gamma^{ij}_{BA} = -
{1 \over 2} E^{B2} \wedge
f^{i}
\Gamma^{i}_{BA}. \qquad
\end{equation}
Here the 'admissible derivatives' of the harmonics \cite{BZ} (i.e. the
derivatives
which preserve
the conditions (\ref{vharm}), (\ref{sharmD0}))
\begin{equation}\label{du0}
du^{~\underline{a}}_{\underline{m}} =
u_{\underline{b}\underline{m}}
\Omega^{\underline{b}\underline{a}} \qquad \Leftrightarrow \qquad
\cases{
du^{(0)}_{\underline{m}} =
u^i_{\underline{m}} f^i, \cr
du^{i}_{\underline{m}} = - u^j_{\underline{m}} A^{ji} +
u^{(0)}_{\underline{m}} f^i, \cr }
\end{equation}
\begin{equation}\label{dv}
dv^{~A}_{\underline{\mu}}
\equiv
{1 \over 4} \Omega^{\underline{b}\underline{a}}
v^{~B}_{\underline{\mu}}
(\sigma_{\underline{b}\underline{a}})_B^{~A} =
{1 \over 2}
v^{~B}_{\underline{\mu}} f^i
\Gamma^i_{BA}
-
{1 \over 4} A^{ij}
v^{~B}_{\underline{\mu}}
\Gamma^{ij}_{BA}
\end{equation}
\begin{equation}\label{dv-1}
dv_{A}^{~\underline{\mu}}
\equiv
- {1 \over 4} \Omega^{\underline{b}\underline{a}}
(\sigma_{\underline{b}\underline{a}})_A^{~B}
v_{B}^{~\underline{\mu}}
=
{1 \over 2}
f^i
\Gamma^i_{AB}
v_{B}^{~\underline{\mu}}
+
{1 \over 4} A^{ij}
\Gamma^{ij}_{AB}
v_{B}^{~\underline{\mu}}
\end{equation}
have been used.
In (\ref{du0}), (\ref{dv}), (\ref{dv-1})
\begin{equation}\label{Cf}
 \Omega^{\underline{a}\underline{b}}(d)
 = -  \Omega^{\underline{b}\underline{a}}(d)
  \equiv
u^{\underline{a}}_{\underline{m}} d u^{\underline{b}\underline{m}}
 = \pmatrix {  0     &  f^{j}  \cr
            - f^{i}  &  A^{ij} \cr}
\end{equation}
are $so(1,9)$--valued Cartan 1--forms. The forms
 \begin{equation}\label{fi}
f^{i} \equiv u^{(0)}_{\underline m} d u^{\underline m~i}
\end{equation}
are covariant with respect to local $SO(9)$ transformations and provide a
basis for the coset $SO(1,9)/SO(9)$ while
\begin{equation}\label{Aij}
A^{ij} \equiv u^{i}_{\underline m} d
u^{\underline m~j}
\end{equation}
 transform as $SO(9)$ connections.
 From the  definition (\ref{Cf}) one can find that the Cartan forms
 satisfy a zero curvature conditions (Maurer--Cartan equations)
\begin{equation}\label{MC}
d\Omega^{\underline{a}~\underline{b}} - \Omega^{\underline{a}\underline{c}}
\wedge
\Omega_{\underline{c}}^{~\underline{b}} = 0 \quad \Leftrightarrow \quad
\cases{
{\cal D} f^i = df^i + f^j \wedge A^{ji} = 0, \cr
F^{ij}= dA^{ij} + A^{ik} \wedge A^{kj} =
- f^i \wedge f^j, \cr}
\end{equation}

 \section{Gauge symmetries and equations of motion for free
 super--D0--brane}

 The simplest way to vary the geometric action is to calculate an external
 derivative of the Lagrangian 1--form
 (\ref{SD0})
\begin{equation}\label{L}
{{\cal L}}_1 =  m \left[E^{(0)} +
i  \left( d{\Theta}^{1\underline{\mu}}
{\Theta}^{2}_{\underline{\mu}} -
{\Theta}^{1\underline{\mu}} d{\Theta}^{2}_{\underline{\mu}}
\right)\right],
\end{equation}
 (cf. (\ref{SD0}),
 (\ref{E0i}), (\ref{EA1,2})) and use the seminal formula
\begin{equation}\label{vL}
\delta {{\cal L}}_1 =
i_\delta
(d{{\cal L}}_1) + d i_\delta
{{\cal L}}_1.
\end{equation}
Here $i_\delta $ can be regarded as formal contraction of differential
form with variation symbol, e.g.
\begin{equation}\label{var}
i_\delta d{\Theta}^{1\underline{\mu}} = \delta
{\Theta}^{1\underline{\mu}},
\quad
i_\delta d{\Theta}^{2}_{\underline{\mu}} =
\delta {\Theta}^{2}_{\underline{\mu}} , \quad
i_\delta {\Pi}^{\underline{m}} =
\delta X^{\underline{m}} -
i \delta {\Theta}^{1}
{\sigma}^{\underline{m}}
{\Theta}^{1}
-
i \delta {\Theta}^2
\tilde{\sigma}^{\underline{m}}
{\Theta}^2.
\end{equation}
The basis (\ref{var}) in the space of variations is more convenient than
the original one
$\left(\delta X^{\underline{m}}, \delta {\Theta}^{1\underline{\mu}},
\delta {\Theta}^{2}_{\underline{\mu}} \right)$.
The contractions $i_\delta f^i, i_\delta A^{ij}$ shall be considered as
parameters of independent transformations of the harmonic variables
which preserve the conditions (\ref{vharm}), (\ref{sharmD0})
(admissible variations \cite{BZ})
 \begin{equation}\label{vu0i}
\delta u^{(0)}_{\underline{m}} =
u^i_{\underline{m}}
i_\delta f^i, \qquad \delta u^{i}_{\underline{m}} = -
u^j_{\underline{m}} i_\delta A^{ji} + u^{(0)}_{\underline{m}}
i_\delta f^i,
\end{equation}
\begin{equation}\label{vv}
\delta v^{~A}_{\underline{\mu}}
= {1 \over 2}
v^{~B}_{\underline{\mu}}
\Gamma^i_{BA}
~i_\delta f^i~
-
{1 \over 4}
v^{~B}_{\underline{\mu}}
\Gamma^{ij}_{BA}
~i_\delta A^{ij}~
, \qquad
\end{equation}
$$
\delta v_{A}^{~\underline{\mu}} =
{1 \over 2}
\Gamma^i_{AB}
v_{B}^{~\underline{\mu}}~
i_\delta f^i ~
+
{1 \over 4}
\Gamma^{ij}_{AB}
v_{B}^{~\underline{\mu}}
~i_\delta A^{ij}~.
$$

External derivative of the Lagrangian 1--form  (\ref{L})
can be written as
\begin{equation}\label{dL}
d{\cal L}_1 = m \left[E^i \wedge f^i - i
(E^{A1} - E^{A2}) \wedge (E^{A1} - E^{A2})\right].
\end{equation}
Thus the variation of the action (\ref{SD0}) is
\begin{equation}\label{vSD0}
S_{D0} = m \int_{{\cal M}^1}
\left(
E^i i_\delta f^i -
f^i i_\delta E^i -
2i
(E^{A1} - E^{A2}) i_\delta  (E^{A1} - E^{A2})
\right),
\end{equation}
where we skipped the complete derivative term
$\int_{{\cal M}^1} d i_\delta {{\cal L}_1}$.
 The latter means that the
D0-brane worldline is considered as a surface without boundary
$\partial {\cal M}^1=0$ and, hence,
there are no rejections for its identification with a boundary
of some surface ${\cal M}^1=\partial {\cal M}^{1+1}$
(see below).

Only $16$ of $32$ fermionic variations
\begin{equation}\label{fv}
i_\delta  (E^{A1} - E^{A2})\equiv
  \delta {\Theta}^{1\underline{\mu}} v^{~A}_{\underline{\mu}} -
  \delta {\Theta}^2_{\underline{\mu}} v_{A}^{~\underline{\mu}}
\end{equation}
are involved effectively
in (\ref{vSD0}).
Thus the remaining $16$ variations
\begin{equation}\label{kA}
\kappa^A \equiv i_\delta  (E^{A1} + E^{A2})\equiv
  \delta {\Theta}^{1\underline{\mu}} v^{~A}_{\underline{\mu}} +
  \delta {\Theta}^2_{\underline{\mu}} v_{A}^{~\underline{\mu}}
\end{equation}
can be regarded as parameters of a fermionic gauge symmetry of the model.
This is the famous $\kappa$--symmetry \cite{AL}\footnote{
See \cite{stv} for the geometrical meaning of the $\kappa$--symmetry.}.
Other gauge symmetries can be found by searching for the variations
whose parameters are absent in (\ref{vSD0}).
They are
$SO(9)$ symmetry ($i_\delta A^{ij}$) and the reparametrization
($i_\delta E^{(0)} =
\delta X^{\underline{m}}
u^{(0)}_{\underline{m}}$, $ \delta \Theta^{1,2}= 0$).

Equations of motion for the super--D0-brane appear as a result of
variations with respect to $ i_\delta f^{i}$, $ i_\delta E^{i} =\delta
X^{\underline{m}} u^{i}_{\underline{m}}$ and $i_\delta (E^{A1}-E^{A2})$
respectively
\begin{equation}\label{Ei=0} \tilde{E}^i \equiv
\tilde{\Pi}^{\underline{m}} \tilde{u}^{i}_{\underline{m}} = 0,
\end{equation}
\begin{equation}\label{fi=0}
\tilde{f}^i \equiv
\tilde{u}^{(0)\underline{m}} d\tilde{u}^{i}_{\underline{m}} = 0,
\end{equation}
\begin{equation}\label{EA-=0}
\tilde{E}^{A1} - \tilde{E}^{A2} \equiv
d\tilde{\Theta}^{1\underline{\mu}} \tilde{v}^{~A}_{\underline{\mu}}
- d\tilde{\Theta}^2_{\underline{\mu}} \tilde{v}_{A}^{~\underline{\mu}}
= 0.
\end{equation}
It can be proved that these equations are equivalent to the
standard equations of motion for the super--D0--brane \cite{bt}.
In the gauge $\tilde{X}^{\underline{m}} \tilde{u}^{(0)}_{\underline{m}} =
\tau $, $\tilde{\Theta}^{1\underline{\mu}} =0$ Eqs. (\ref{Ei=0}),
 (\ref{fi=0}), (\ref{EA-=0}) are equivalent to the set
\begin{equation}\label{eqg}
d\tilde{X}^{\underline{m}}= d\tau  p^{\underline{m}}/m, \quad
dp_{\underline{m}}=0, \quad
p^2_{\underline{m}}=m^2, \quad
d\tilde{\Theta}^2_{\underline{\mu}}=0,
\end{equation}
 which describes a massive superparticle.

\section{Geometric action for type $IIA$ superstring}

The geometric action, superembedding approach  and generalized action
principle for type $IIB$ superstring has been constructed in
\cite{BZ,bpstv,bsv} respectively. Type $IIA$ superstring has not
been considered in this framework before.

The geometric action for type IIA superstring is
  \begin{equation}\label{SIIA}
S_{IIA}=  \int_{{\cal M}^{1+1}}
{\hat{\cal L}}_2 =
\int_{{\cal M}^{1+1}}
\left( {1 \over 2} \hat{E}^{++} \wedge  \hat{E}^{--} - \hat{B}_2
\right),
\end{equation}
where
\begin{equation}\label{EpmI}
 E^{\pm\pm} \equiv \Pi^{\underline{m}} U^{\pm\pm }_{\underline{m}}, \qquad
 E^I \equiv \Pi^{\underline{m}} U^{I}_{\underline{m}} ,
\end{equation}
\begin{equation}\label{B2}
 B_2 = i \Pi^{\underline{m}}
 \wedge \left(
d{\Theta}^{1}
{\sigma}_{\underline{m}}
{\Theta}^{1} -
i d{\Theta}^2
\tilde{\sigma}_{\underline{m}}
{\Theta}^2\right)
 +
d{\Theta}^{1}
{\sigma}^{\underline{m}}
{\Theta}^{1} \wedge
d{\Theta}^2
\tilde{\sigma}_{\underline{m}}
{\Theta}^2,
\end{equation}
\begin{equation}\label{hatPi}
\hat{\Pi}^{\underline{m}} =
d\hat{X}^{\underline{m}} -
i d\hat{\Theta}^{1}
{\sigma}^{\underline{m}}
\hat{\Theta}^1
-
i d\hat{\Theta}^2
\tilde{\sigma}^{\underline{m}}
\hat{\Theta}^2 = d\xi^m \hat{\Pi}^{~\underline{m}}_m (\xi )
\end{equation}
is the pull-back of the 1--form (\ref{Pi}) on the superstring
worldsheet
${\cal M}^{1+1}$ whose embedding
into the type $IIA$ superspace
${\cal M}^{1+1} \rightarrow {\underline{{\cal
 M}}}^{(10|32)}$ is defined by
\begin{equation}\label{M1+1}
{X}^{\underline{m}} =
\hat{X}^{\underline{m}} (\xi)
\equiv
\hat{X}^{\underline{m}} (\tau, \sigma ), \quad
{\Theta}^{1\underline{\mu}} =
\hat{\Theta}^{1\underline{\mu}}
 (\xi), \quad
{\Theta}^2_{\underline{\mu}} =
\hat{\Theta}^2_{\underline{\mu}}
 (\xi). \quad
\end{equation}
$\hat{U}^{\pm\pm}_{\underline{m}} (\xi)
\equiv
\hat{U}^{0}_{\underline{m}} (\xi) \pm
\hat{U}^{9}_{\underline{m}} (\xi)$
are light--like Lorentz harmonic vectors \cite{Sok}, i.e. the
components  of the $SO(1,9)$ valued matrix
\begin{equation}\label{vhIIA}
U^{~\underline{a}}_{\underline{m}}
=
\left( U^{0}_{\underline{m}},
U^{~I}_{\underline{m}},
U^{9}_{\underline{m}}\right)
=
\left(
{1 \over 2} (U^{++}_{\underline{m}}+U^{--}_{\underline{m}}),
U^{~I}_{\underline{m}},
{1 \over 2} (U^{++}_{\underline{m}}-U^{--}_{\underline{m}})
\right)
~ \in ~ SO(1,9)
\end{equation}
The spinor harmonics
\begin{equation}\label{shIIA}
V^{~\underline{\alpha} }_{\underline{\mu}}
= \left(
V^{~+}_{\underline{\mu}q},
V^{~-}_{\underline{\mu}\dot{q}}\right)^T
\quad \in \quad Spin(1,9)
\end{equation}
\begin{equation}\label{sh-1IIA}
V_{\underline{\alpha} }^{~\underline{\mu}}
= \left(
V^{-\underline{\mu}}_q,
V^{+\underline{\mu}}_{\dot{q}}\right)
\quad \in \quad Spin(1,9)
\end{equation}
\begin{equation}\label{vv-1IIA}
V_{\underline{\alpha} }^{\underline{\mu}}
V^{~\underline{\beta} }_{\underline{\mu}}
=
\delta^{~\underline{\beta} }_{\underline{\alpha}}: \quad
V^{-\underline{\mu}}_q V^{~+}_{\underline{\mu}p}= \delta_{qp}, ~~
V^{+\underline{\mu}}_{\dot{q}}
V^{~-}_{\underline{\mu}\dot{p}}=
\delta_{\dot{q}\dot{p}}, ~~
V^{-\underline{\mu}}_q
V^{~-}_{\underline{\mu}\dot{p}}=
V^{+\underline{\mu}}_{\dot{p}}
V^{~+}_{\underline{\mu} q}=0
\end{equation}
 are related with (\ref{vhIIA}) by Eqs. (\ref{vss}) which include,
 in particular,
\begin{equation}\label{vssA1}
 U^{++}_{\underline{m}}
\sigma^{\underline{m}}_{\underline{\mu}\underline{\nu}}
= 2V_{\underline{\mu}q}^{~+}
V^{~+}_{\underline{\nu}q}, \qquad
 U^{--}_{\underline{m}}
\tilde{\sigma}^{\underline{m}\underline{\mu}\underline{\nu}}
= 2V^{-\underline{\mu}}_{q}
V_{q}^{-\underline{\nu}}, \qquad
\end{equation}
\begin{equation}\label{vssA2}
 U^{--}_{\underline{m}}
\sigma^{\underline{m}}_{\underline{\mu}\underline{\nu}}
= 2V_{\underline{\mu}\dot{q}}^{~-}
V^{~-}_{\underline{\nu}\dot{q}}, \qquad
 U^{++}_{\underline{m}}
\tilde{\sigma}^{\underline{m}\underline{\mu}\underline{\nu}}
= 2V^{+\underline{\mu}}_{\dot{q}}
V_{\dot{q}}^{+\underline{\nu}}. \qquad
\end{equation}
The details about the stringy harmonics and Cartan forms
(cf. (\ref{Cf}))
\begin{equation}\label{Cfst}
 \Omega^{\underline{a}\underline{b}}
  \equiv
U^{\underline{a}}_{\underline{m}} d U^{\underline{b}\underline{m}}
 = \left( \matrix{
 0 & { f^{++J} + f^{--J}\over 2} & -{1 \over 2}\omega   \cr
-{ f^{++I} + f^{--I}\over 2} &  A^{IJ} & - { f^{++I} - f^{--I}\over 2} \cr
  {1 \over 2}\omega  &  { f^{++J} - f^{--J}\over 2}  & 0 \cr }  \right)
\end{equation}
can be found in Refs. \cite{BZ,bpstv,bsv,BK}.

The external derivative of the Lagrangian 2--form
  \begin{equation}\label{L2}
{{\cal L}}_2 =
 {1 \over 2} {E}^{++} \wedge  {E}^{--} -
 i \Pi^{\underline{m}}
 \wedge \left(
d{\Theta}^{1}
{\sigma}_{\underline{m}}
{\Theta}^{1} -
i d{\Theta}^2
\tilde{\sigma}_{\underline{m}}
{\Theta}^2\right)
 +
d{\Theta}^{1}
{\sigma}^{\underline{m}}
{\Theta}^{1} \wedge
d{\Theta}^2
\tilde{\sigma}_{\underline{m}}
{\Theta}^2
\end{equation}
can be calculated with the use of
(\ref{vssA1}),
(\ref{vssA2}), (\ref{Cfst}) and stringy counterparts of
Eqs.
(\ref{du0})--
(\ref{MC})
\begin{equation}\label{dL2}
d{{\cal L}}_2 =
-2i {E}^{++}
\wedge  {E}^{-\dot{q}1}
\wedge  {E}^{-\dot{q}1}
+2i {E}^{--}
\wedge  {E}^{+2}_{\dot{q}}
\wedge  {E}^{+2}_{\dot{q}}  +
\end{equation}
 $$
E^I \wedge \left(
{1\over 2}
E^{--} \wedge f^{++I} -
{1\over 2}
E^{++} \wedge f^{--I} +
2i \left( E^{+q1} \wedge E^{-\dot{q}1} +
E^{+2}_{q} \wedge E^{-2}_{\dot{q}} \right) \gamma^I_{q\dot{q}}
 \right).
$$
The parameters of the stringy $\kappa$--symmetry can be identified
with the contractions of those fermionic forms which are absent
in the first line of Eq. (\ref{dL2})
\begin{equation}\label{kIIA}
\kappa^{+q}
 \equiv i_\delta E^{+q1} =
  \delta {\Theta}^{1\underline{\mu}} v^{~+}_{\underline{\mu}q}, \qquad
\kappa^{-{q}}
\equiv i_\delta E^{-2}_{{q}} =
  \delta {\Theta}^2_{\underline{\mu}} v_{q}^{-\underline{\mu}}.
\end{equation}
The second line of (\ref{dL2}) determines, in particular, the
transformations of the harmonics with respect to the $\kappa$--symmetry.
Other gauge symmetries are $SO(1,1) \times SO(8)$ ($i_\delta \omega , ~
i_\delta A^{IJ}$) and the reparametrization
($ i_\delta E^{\pm\pm} = \delta X^{\underline{m}}
U^{\pm\pm}_{\underline{m}} $).

The equations of motion for the type $IIA$ superstring can be obtained
from (\ref{dL2}). They are
\begin{equation}\label{EI}
\hat{E}^{I} \equiv
  \hat{\Pi}^{\underline{m}} u^{I}_{\underline{m}}=0, \qquad
\end{equation}
\begin{equation}\label{fI}
M^I_2 \equiv E^{--} \wedge f^{++I} - E^{++} \wedge f^{--I}
+ 4i \left( E^{+q1} \wedge E^{-\dot{q}1} +
E^{+2}_{q} \wedge E^{-2}_{\dot{q}} \right) \gamma^I_{q\dot{q}} = 0,
\end{equation}

\begin{equation}\label{dTh1}
{E}^{++} \wedge  {E}^{-\dot{q}1} \equiv
  \hat{\Pi}^{\underline{m}} \wedge
  d{\Theta}^{1\underline{\mu}}
  v^{~-}_{\underline{\mu}\dot{q}} u^{++}_{\underline{m}}=0, \qquad
\end{equation}
\begin{equation}\label{dTh2}
{E}^{--} \wedge  {E}^{+2}_{\dot{q}} \equiv
  \hat{\Pi}^{\underline{m}} \wedge
  d{\Theta}^2_{\underline{\mu}}
  v^{+\underline{\mu}}_{\dot{q}} u^{--}_{\underline{m}}=0. \qquad
\end{equation}
 It can be proved that this set is equivalent to the
 standard equations of motion for the type $IIA$ superstring
 (see \cite{BZ}
 for the type $IIB$ case).

 \section{Supersymmetric action functional for
 type $IIA$ superstring with super--D0--branes at the endpoints}

The main problem which should be solved to write down the action of
interacting branes is: how to take into account an identification of the
bosonic and fermionic superembedding functions on the intersection
\cite{BK,BKD9}.  However, this problem has the natural solution just for
the system under consideration.
Here  the  super--D0--brane  worldline ${\cal M}^{1}$ should be considered
as the boundary
of the superstring worldsheet $~~~{\cal M}^{1} =
\partial  {\cal M}^{1+1}$.  Thus one can define {\sl an embedding}   of
${\cal M}^{1}$ into  ${\cal M}^{1+1}$
\begin{equation}\label{M1dM2}
\xi^m = \tilde{\xi}^m (\tau ) : \quad
{\cal M}^{1} = \partial  {\cal M}^{1+1} \quad \rightarrow {\cal M}^{1+1}
\end{equation}
and identify the super--D0--brane coordinate  functions
$\tilde{X}(\tau ) , \tilde{\Theta}^{1,2}(\tau )$ with the images of the
superstring coordinate functions
$\hat{X} (\xi ), \hat{\Theta}^{1,2}(\xi )$
on the boundary
\begin{equation}\label{idf}
\tilde{X}(\tau ) = \hat{X} \left( \tilde{\xi} (\tau ) \right), \qquad
\tilde{\Theta}^{1,2}(\tau )=
\hat{\Theta}^{1,2}
\left( \tilde{\xi} (\tau ) \right).
\end{equation}
With this identification the action for the coupled system
of an open fundamental superstring and super--D0--branes at the ends of
the superstring is the direct sum of the actions (\ref{SD0}) and
(\ref{SIIA})
\begin{equation}\label{Sint1}
S_{str+D0}=  \int_{{\cal M}^{1+1}} {\hat{\cal L}}_2 + m \int_{\partial
{\cal M}^{1+1}} {\tilde{\cal L}}_1.
\end{equation}
The variation of the action can be calculated as
\begin{equation}\label{dSint1} \delta
S_{str+D0}=  \int_{{\cal M}^{1+1}} i_\delta \left(d{\cal L}_2 \right) + m
\int_{\partial {\cal M}^{1+1}} \left( i_\delta {\cal L}_2 + i_\delta
d{{\cal L}}_1 \right).
\end{equation}
A possibility is to require the vanishing of the bulk and the boundary
variations in  (\ref{dSint1}) separately.
On the other hand, following \cite{BK,BK2,BKD9}, one can introduce the
following current density distribution
\begin{equation}\label{j1} j_1 = d\xi^{m} \varepsilon_{mn}
\int_{\partial {\cal M}^{2}} d{\tilde \xi}^{n} (\tau ) \delta^{2} \left(
\xi - {\tilde \xi} (\tau )\right)~= \varepsilon_{mn} d\xi^{n} j^{m}
\qquad
\end{equation}
with the property
\begin{equation}\label{intj1}
\int_{{\cal M}^{1+1}} j_1 \wedge
{\hat {\cal A}}_1  =
\int_{\partial {\cal M}^{1+1}} {\tilde {\cal A}}_1.
\end{equation}
In (\ref{intj1}) ${\hat {\cal A}}_1$  is an arbitrary  1-form defined on
the worldsheet ${\cal M}^{1+1}$ and ${\tilde{\cal A}}_1$ is its pull--back
onto ${\cal M}^{1} = \partial {\cal M}^{1+1}$.
As it is easy to define the extension of the super--D0--brane
Lagrangian form to the whole worldsheet (\ref{L}), we can use (\ref{intj1})
to lift the action (\ref{Sint1}) or its variation (\ref{dSint1}) to the
integral over the whole worldsheet
\begin{equation}\label{Sint}
S_{str+D0}= \int_{{\cal M}^{1+1}} {\hat{\cal L}}_2 + m j_1 \wedge
{\hat{\cal L}}_1,
\end{equation}
\begin{equation}\label{dSint2} \delta
S_{str+D0}=  \int_{{\cal M}^{1+1}} i_\delta \left(d{\cal L}_2 \right) +
m j_1 \wedge \left( i_\delta {\cal L}_2 + i_\delta d{{\cal L}}_1 \right).
\end{equation}
Here ${\hat{\cal L}}_2$ is defined by Eq. (\ref{SIIA}), (\ref{L2}),
${\hat{\cal L}}_1$ is the pull--back of the 1--form
${{\cal L}}_1$ (\ref{L}) on the worldsheet
(\ref{M1+1}). In
(\ref{dSint2}) the contractions of the forms
(i.e.
$
i_\delta {\cal B}_2 =
i_\delta (1/2 dZ^M \wedge dZ^N {\cal B}_{NM}) =
 dZ^{\underline{M}} \delta Z^{\underline{N}}
 {\cal B}_{\underline{N}\underline{M}}$ )
 should be pulled back to the worldsheet and the
 contractions of the coordinate differentials in the second
 term
 should include the variations of the functions $\tilde{\xi }^m (\tau )$
 (\ref{M1dM2}), e.g. $i_\delta d{X}^{\underline{m}} = \delta
 \tilde{X}^{\underline{m}} + \delta \tilde{\xi }^m (\tau )
 \partial_m \hat{X}^{\underline{m}}\vert_{\xi = \tilde{\xi}(\tau)}$
 (though the variations $\delta \tilde{\xi }^m (\tau )$ shall not
 produce independent equations).

\section{D0-branes at the endpoints of bosonic string}

In the pure bosonic limit our dynamical  system
describes a 'string with masses at the endpoints'.
Such system has been studied in early years of QCD strings and
 partial solutions
 have been found \cite{BarNes}.
 However the geometric or first order formulation as well as
 'extended variational problem' approach \cite{BK2}
 (\ref{Sint}),  (\ref{dSint2})  for this system
 are new and, in our opinion,
 instructive.

 The geometric action for the coupled system has the form
\begin{equation}\label{Sintb}
S =
\int_{{\cal M}^{1+1}} {1 \over 2} \hat{E}^{++} \wedge \hat{E}^{--} +
m j_1 \wedge \hat{E}^{(0)}.
\end{equation}
Its variation with respect to harmonic variables
\begin{equation}\label{vhSb}
\delta_{h}S =
\int_{{\cal M}^{1+1}} \left(
{1 \over 2} \hat{E}^{I} \wedge \hat{E}^{--} i_\delta f^{++I} -
{1 \over 2} \hat{E}^{I} \wedge \hat{E}^{--} i_\delta f^{++I} +
m j_1 \wedge \hat{E}^{i} i_\delta f^{i}\right)
\end{equation}
produces the same algebraic embedding equations
as in the case of a free string and free D0-brane(s)
\begin{equation}\label{EIb}
\hat{E}^{I} \equiv
  d\hat{X}^{\underline{m}}(\xi) U^{I}_{\underline{m}}(\xi) = 0, \qquad
\end{equation}
\begin{equation}\label{Eib}
\tilde{E}^{i} \equiv
  d\hat{X}^{\underline{m}}(\xi (\tau)) u^{i}_{\underline{m}}(\tau)=0.
  \qquad \end{equation}
  This provides the possibility to simplify the
variation with respect to the coordinate functions considering it modulo
Eqs.  (\ref{EIb}), (\ref{Eib})
\begin{equation}\label{vSb}
\delta S
\vert_{\hat{E}^I=0=\tilde{E}^i} =  {1 \over 2} \int_{{\cal M}^{1+1}}
\left( \hat{M}_2^I U^I_{\underline{m}} + j_1 \wedge \left( \hat{E}^{++}
U^{--}_{\underline{m}} - \hat{E}^{--} U^{++}_{\underline{m}} - 2 m f^i
u^{i}_{\underline{m}} \right) \right) \delta \hat{X}^{\underline{m}}
\end{equation}
Here
\begin{equation}\label{MI}
M^I_2 \equiv E^{--} \wedge f^{++I} - E^{++} \wedge f^{--I}
\end{equation}
is the pure bosonic limit of the l.h.p. of the free superstring equation
(\ref{fI}).

When considered together, Eqs. (\ref{EIb}) and (\ref{Eib}) relate the
images
of the stringy harmonics on the worldsheet boundary with the harmonics of
the D0-branes. Indeed, Eq. (\ref{Eib}) implies
$d\tilde{X}^{\underline{m}}= E^{(0)} u^{(0) \underline{m}}$. Substituting
this relation into the pull-back of Eq. (\ref{EIb}) on the boundary
one arrives at
\\ $ u^{(0)\underline{m}}(\tau)~U^{I}_{\underline{m}}(\xi(\tau ))=0$. This
implies
\begin{equation}\label{uUU}
u^{(0) \underline{m}}(\tau) =
w^{--}(\tau ) U^{++ \underline{m}} (\xi (\tau ))+
U^{++ \underline{m}} (\xi
(\tau ))/ 4 w^{--}(\tau )
\end{equation}
with some indefinite function $w^{--}(\tau )$ (compensator for
$SO(1,1)$ symmetry).
The relative coefficient in (\ref{uUU}) is fixed by normalization
conditions.
Now, using the $SO(9)$ gauge symmetry,
we can chose the super--D0--brane harmonics to be expressed through
the images of stringy ones by
\begin{equation}\label{u=U}
u^{i\underline{m}}(\tau) = \left(
U^{++ \underline{m}} (\xi (\tau )),
u^{(9)\underline{m}}(\tau) \right),
\end{equation}
$$ u^{(9)\underline{m}}(\tau) = w^{--}(\tau )
U^{++ \underline{m}} (\xi (\tau ))-
U^{++ \underline{m}} (\xi (\tau ))/ 4 w^{--}(\tau ).
$$
Then the set of Cartan forms (\ref{fi}) splits as
$f^{i}=(f^{I}, f^{(9)})$ and, after some algebraic
manipulations, the equations of
 motion can be written in the form
 \begin{equation}\label{MI=} M^I_2
 \equiv E^{--} \wedge f^{++I} - E^{++} \wedge f^{--I} = m j_1 \wedge
 f^{I},
 \end{equation}
 \begin{equation}\label{j1f9}
j_1 \wedge \left(
 E^{(0)}- m f^{(9)}
 \right)=0.
 \end{equation}
  When $m\not= 0$ the latter equation evidently implies
 \begin{equation}\label{f9=}
 f^{(9)} \equiv u^{(0)\underline{m}}d u^{(9)}_{\underline{m}} =
 {1 \over m} E^{(0)}.
 \end{equation}
 For $m\rightarrow \infty$ we can neglect the left hand side of
 Eq. (\ref{MI=}) and the right hand side of Eq.(\ref{f9=}). Thus
 we arrive at the free  equations of motion (\ref{fi=0})
 for the D0-branes.
 This means that D0--branes (or 'quarks' \cite{BarNes}) with infinite
 mass(es)
 do not feel the influence of the open string.  When $m\rightarrow 0$
Eq. (\ref{MI=})  becomes the free string equation, while  (\ref{j1f9})
implies that $j_1=0$, i.e.  that the worldsheet has no boundary and,
thus, the string is closed.

\section*{Concluding Remarks}

The analyzes of the gauge symmetries of the action
(\ref{Sint1}), (\ref{Sint}) and the supersymmetric equations which
follow from it will be the subject of a forthcoming article.
We expect that they shall provide an important insights for
future study of the generic system of interacting superbranes.

Another direction of the development of the present results
is to elaborate the generalized action principle
\cite{bsv} and  the superembedding approach \cite{bpstv,bsv,hsw}
the super--D0--brane
(see \cite{bst} for the super-Dp-branes with
$0<p<9$ and \cite{abkz} for $p=9$).
The basis for such study is provided by the geometric action (\ref{SD0}).

\section*{Acknowledgments}
The author is grateful to D. Sorokin, M. Tonin, B. Julia
for useful conversations and for the hospitality at the
Padova Section of INFN (Padova) and Laboratoire de Physique Theorique de
l'Ecole Normale Superieure (Paris), where
a part of this work has been done.  A partial support from the INTAS Grant
{\bf 96-308} and the Ukrainian GKNT Grant {\bf 2.5.1/52} is acknowledged.

\end{document}